\documentclass[twocolumn,showpacs,preprintnumbers,prl]{revtex4}
\usepackage{amssymb}
\usepackage{amsmath}
\usepackage{graphicx}
\usepackage{dcolumn}
\usepackage{bm}

\begin{document}

\title{Measurable Entanglement for Tripartite Quantum Pure States of Qubits}
\author{Chang-shui Yu}
\author{He-shan Song}
\email{hssong@dlut.edu.cn}
\affiliation{Department of Physics, Dalian University of Technology,\\
Dalian 116024, P. R. China}
\date{\today }

\begin{abstract}
We show that for tripartite quantum pure states of qubits, all the
kinds of entanglement in terms of SLOCC classification are
experimentally measurable by simple projective measurements,
provided that four copies of the composite quantum system are
available. In particular, the entanglement of reduced density
matrices, even though they are mixed states, can be exactly
determined in experiment. Concurrence of assistance is also shown to
be measurable by introducing an interesting equations with explicit
physical meanings.
\end{abstract}

\maketitle

\emph{Introduction.}- Quantum entanglement is not only one of the
fundamental characteristics that distinguishes the quantum from the
classical world, but also an important physical resource for quantum
information processing. A lot of measures have been presented to
quantify entanglement [1]. However, up till now, no directly
measurable observable corresponds to entanglement of a given
arbitrary quantum state, owing to the unphysical quantum operations
in usual entanglement measure [2], for example, the complex
conjugation of concurrence [3] and the partial transpose of
negativity [4,5]. To evaluate the entanglement in experiment, a
general approach is to reconstruct the density matrix by measuring a
complete set of observables [6-8], which is only suitable for small
quantum systems . Entanglement witnesses have been proven effect for
the detection of entanglement [9], however they depend on the
detected states, which implies that a priori knowledge on the states
is required. Quite recently, some approaches have been reported for
the determination of entanglement in experiment [2,10-12]. The most
remarkable one is the new formulation of concurrence [13] in terms
of copies of the state which led to the first direct experimental
evaluation of entanglement [12]. Later, a measurable multipartite
concurrence in terms of a single factorizable observable was
presented [14]. The concurrence in terms of copies of states was
generalized to mixed states [15], which in fact provides an
observable lower bound of concurrence of mixed states and could be
understood as a generalized entanglement witness. A natural problem
is whether the tripartite entanglement is experimentally measurable.

Unlike bipartite entanglement which can be quantified by only a
single quantity due to that any state can be prepared from a
maximally entangled state by means of local operations and classical
communication (LOCC), in general, tripartite entanglement can not be
effectively quantified by a single scalar quantity because three
qubits can be entangled in different ways [16]. It is obvious that
different kinds of entanglement of a tripartite pure states can not
be experimentally determined by the expectation of a single
observable. In this Letter, we show that for an arbitrary tripartite
quantum pure state of qubits it is possible to directly measure all
the different kinds of entanglement based on four different
projective measurements, provided that four copies of the tripartite
quantum pure state are available. In particular, even the reduced
density matrices are mixed, the exact entanglement instead of lower
bound can be experimentally determined. In addition, an interesting
equation with explicit physical meanings has been introduced by
which we show that concurrence of assistance (COA)[17,18] is
measurable.

\emph{Description of entanglement of tripartite quantum pure states
of qubits.}- A tripartite quantum pure states of qubits defined in
the Hilbert space $\mathcal{H}=\mathcal{H}_1\otimes
\mathcal{H}_2\otimes \mathcal{H}_3$ can be written in standard basis
by
\begin{equation*}
\left\vert
\psi\right\rangle_{ABC}=\sum_{i,j,k=0}^{1}a_{ijk}\left\vert
i\right\rangle_A\left\vert j\right\rangle_B\left\vert k\right\rangle_C.%
\eqno{(1)}
\end{equation*}
It can be divided into six inequivalent classes under stochastic
local operations and classical communication (SLOCC) [16], i.e.
(i)unentangled states (tripartite separable states), if $\left\vert
\psi\right\rangle_{ABC}=\left\vert
\phi\right\rangle_A\otimes\left\vert\chi\right\rangle_B\otimes\left\vert\eta%
\right\rangle_C$; (ii)A-to-(BC) bipartite separable states, if $\left\vert
\psi\right\rangle_{ABC}=\left\vert
\phi\right\rangle_A\otimes\left\vert\varphi\right\rangle_{BC}$;
(iii)B-to-(AC) bipartite separable states, if $\left\vert
\psi\right\rangle_{ABC}=\left\vert
\phi\right\rangle_B\otimes\left\vert\varphi\right\rangle_{AC}$;
(iv)C-to-(AB) bipartite separable states, if $\left\vert
\psi\right\rangle_{ABC}=\left\vert
\phi\right\rangle_{AB}\otimes\left\vert\varphi\right\rangle_{C}$; (v)
GHZ-type genuine tripartite entangled states with the standard form
\begin{equation*}
\left\vert GHZ\right\rangle=\frac{1}{\sqrt{2}}\left (\left\vert
000\right\rangle+\left\vert 111\right\rangle\right),\eqno{(2)}
\end{equation*}
and (vi) W-type genuine tripartite entangled states with the standard form
\begin{equation*}
\left\vert W\right\rangle=\frac{1}{\sqrt{3}}\left
(\left\vert
001\right\rangle+\left\vert 010\right\rangle+\left\vert
100\right\rangle\right).\eqno{(3)}
\end{equation*}

A direct and complete description of entanglement of tripartite
quantum pure states of qubits is to define a four dimensional vector
named entanglement vector which can be used to distinguish and
quantify all the
kinds of entanglement. For instance, define%
\begin{equation*}
\mathcal{E}(\left\vert \psi\right\rangle_{ABC})=\left[%
E_{ii},E_{iii},E_{iv},E_{v}\right],\eqno{(4)}
\end{equation*}
where $E_{ii}=C(\left\vert\psi\right\rangle_{A(BC)})$ denotes
A-to-(BC) bipartite concurrence; $E_{iii}$ and $E_{iv}$ corresponds
to B-to-(AC) and C-to-(AB) bipartite concurrence, respectively;
$E_{v}$ denotes the 3-tangle introduced in Ref. [19]. It is obvious
that $\mathcal{E}=0$ corresponds to
class (i); $E_{m}=0,m=ii,iii,iv$ corresponds to the \emph{m}th class. $%
E_{v}\neq 0$ shows the existence of GHZ-type entanglement. Furthermore, let $%
\rho_{AB}$ denote the reduced density matrix of two qubits,
following the remarkable Coffman-Kundu-Wootters equation [19]
\begin{equation*}
E_{v}+C^2(\rho_{AB})+C^2(\rho_{AC})=C^2(\left\vert\psi\right\rangle_{A(BC)}),%
\eqno{(5)}
\end{equation*}
and those corresponding to other foci, one can always determine the
entanglement of reduced density matrices in terms of $\mathcal{E}$.
Thus the existence of W-type entanglement can also be determined
because the W-type relevant entanglement measure can always be given
in terms of the entanglement of reduced density matrices [20]. In a
word, so long as the entanglement vector $\mathcal{E}$ is given, all
the kinds of entanglement can be determined. That is to say, if
$\mathcal{E}$ is measurable, all the entanglement including those of
the mixed reduced density matrices can be exactly determined in
experiment. Ref. [14] has implied that the bipartite concurrence
such as $C(\left\vert\psi\right\rangle_{A(BC)})$ is measurable by a
simple projective measurement if two copies of the state are
available, therefore all the remaining are to prove 3-tangle can
also be measurable.

\emph{Measurable 3-tangle.}-3-tangle can be defined by
\begin{equation*}
\tau (\left\vert \psi \right\rangle _{ABC})=4\left\vert\det
{R}\right\vert,\eqno{(6)}
\end{equation*}%
where
\begin{equation*}
R_{ij}=\left\langle \psi ^{\ast }\right\vert _{ABC}\left(\sigma
_{y}\otimes \sigma _{y}\otimes \left\vert i\right\rangle
\left\langle j\right\vert \right)\left\vert \psi \right\rangle
_{ABC},\eqno{(7)}
\end{equation*}%
with $i,j=0,1$. $\sigma _{y}=\left(
\begin{array}{cc}
0 & -i \\
i & 0%
\end{array}%
\right) $, and $\{\left\vert i\right\rangle \}$ denotes the magic
basis of $\mathcal{H}_{3}$. Consider the fourfold copy
$\otimes_{k=1}^4\left\vert\psi\right\rangle_{ABC}$ of
$\left\vert\psi\right\rangle_{ABC}$, one can define
$$P_-^{(i_mi_n)}=\frac{1}{\sqrt{2}}\left(\left\vert 0\right\rangle_{i_m}
\left\vert1\right\rangle_{i_n} -\left\vert 1\right\rangle_{i_m}\left\vert 0\right\rangle_{i_n}\right), \eqno{(8)}$$
 denoting the projector onto the anti-symmetric
subspace $\mathcal{H}_{i_m}\wedge\mathcal{H}_{i_n}$ of
$\mathcal{H}_{i_m}\otimes\mathcal{H}_{i_n}$ where $i=A,B,C$
corresponds to the subsystems, and $m,n=1,2,3,4$ marks the different
copies of $\left\vert\psi\right\rangle_{ABC}$. Thus a novel
definition of 3-tangle can be derived through the expectation value
of a self-adjoint operator $\mathcal{A}$ as
$$\tau\left(\left\vert\psi\right\rangle_{ABC}\right)=\sqrt{256\left(\otimes_{k=1}^4\left\langle
\psi\right\vert_{ABC}\right)\mathcal{A}\left(\otimes_{k=1}^4\left\vert\psi\right\rangle_{ABC}\right)},\eqno{(9)}
$$
where $\mathcal{A}$ can be formally written by
$$\mathcal{A}=\left[\underset{{j=1,3}}{\underset{k=A,B}{\otimes}}P_-^{(k_jk_{j+1})}\right]\otimes P_-^{(C_1C_3)}\otimes P_-^{(C_2C_4)}.\eqno{(10)}$$
$\mathcal{A}$ is obviously a single factorizable observable. Hence,
$\tau\left(\left\vert\psi\right\rangle_{ABC}\right)$ can be directly
measured in experiment through projective measurements of the
antisymmetric component of the twofold copy
$\left\vert\psi\right\rangle_{ABC}\otimes
\left\vert\psi\right\rangle_{ABC}$ among the four copies. An
illustration of the projective measurements is depicted in Fig. 1.
The analogous projective measurement has been demonstrated for twin
photons in experiment [12].
\begin{figure}[tbp]
\includegraphics[width=8.5cm]{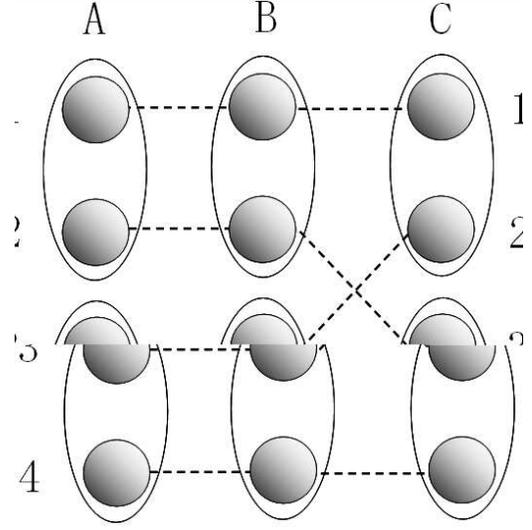}
\caption{The illustration of projective measurements for 3-tangle.
The dotted line denotes the existence of quantum correlation. Every
three balls connected by dotted lines denotes three particles in a
copy of $\left\vert\psi\right\rangle_{ABC}$ with the number 1 to 4
marking the different copies. A, B and C on the top show that the
three particles in a column correspond to the same subsystem. Every
particle can be uniquely denoted by a letter and a number, for
example $A_1$. Every loop represents a projective measurement $P_-$
performed on the two particles inside. Even though a fourfold copy
is necessary for 3-tangle, the projective measurements are only
performed on a twofold copy.} \label{1}
\end{figure}

\emph{Measurable concurrence of assistance.}-Besides the
entanglement classified under SLOCC mentioned above, there are
another two important entanglement measures for tripartite quantum
states, as far as we know. One is the global entanglement which is
defined the same to tripartite concurrence in Ref. [14] and in fact
turned out to be measurable in Ref. [14], the other is the
concurrence of assistance which will be shown to be measurable next
by introducing an interesting equation [21].

For $\left\vert\psi\right\rangle_{ABC}$, COA can be defined [17,18]
by
$$C_a^{(AB)}(\left\vert\psi\right\rangle_{ABC})=Tr\sqrt{\sqrt{\rho_{AB}}\tilde{\rho}_{AB}\sqrt{\rho_{AB}}},\eqno{(11)}$$
with
$\tilde{\rho}_{AB}=(\sigma_y\otimes\sigma_y)\rho_{AB}^*(\sigma_y\otimes\sigma_y)$.
$C_a^{(AB)}(\left\vert\psi\right\rangle_{ABC})$ maximizes the
average concurrence shared by A and B with the help of C. Let
$C(\rho_{AB})$ denote the concurrence of the reduced density matrix
$\rho_{AB}$, then what is the difference between $C_a^{(AB)}$ and
$C(\rho_{AB})$? Let $\lambda_1$ and $\lambda_2$ be the square roots
of the two eigenvalues of $\rho_{AB}\tilde{\rho}_{AB}$, then
$C_a^{(AB)}$ can be rewritten by $\lambda_1+\lambda_2$ and
$C(\rho_{AB})$ can be given by $\left
\vert\lambda_1-\lambda_2\right\vert$. Thus the difference between
them can be directly written by
$$\left[C_a^{(AB)}
\right]^2-C^2(\rho_{AB})=4\lambda_1\lambda_2=\tau\left(\left\vert\psi\right\rangle_{ABC}\right).\eqno{(12)}$$
Consider the concurrence shared by two different parties among A, B
and C, there exist another two analogous equations to eq. (12).
Since both $C(\rho_{AB})$ and
$\tau\left(\left\vert\psi\right\rangle_{ABC}\right)$ can be
experimentally determined, COA is also measurable.

 In fact, besides the main result that eq. (12) shows the measurable COA, eq.
 (12) also has explicit physical meanings. As we know, $C^2(\rho_{AB})$ denotes the
 entanglement of Parties A and B, and $\left[C_a^{(AB)}
\right]^2$ is the maximal average entanglement shared by A and B
with the help of C taken into account. Eq. (12) implies i) COA
includes two contributions: concurrence of the two considered qubits
and three-way entanglement; ii) The role of C is to convert the
three-way entanglement shared by three parties into bipartite
entanglement shared by two parties, thus entanglement shared by two
parties is increased. Two most obvious examples are GHZ state and W
state. The entanglement of reduced density matrix of GHZ state is
zero, hence the COA of GHZ state all comes from the three-way
entanglement and equals to 1 (the value of 3-tangle). On the
contrary, the W state has no three-way entanglement (only two-way
entanglement) [22], hence its COA is only equal to the concurrence
($\frac{4}{9}$) of the two parties. That is to say, for W state,
party C can not provide any help to increase the entanglement
between A and B.

\emph{An alternative description of entanglement of tripartite pure
states.}-One can find any four quantities will be valid for the
entanglement vector if the four quantities can effectively
distinguish and quantify all the kinds of entanglement of tripartite
pure states. From the previous choice of the entanglement vector, it
is not difficult to see that the entanglement vector must be
completely determined in order to evaluate the entanglement of a
single two-qubit reduced density matrix or a single COA. Hence, a
more convenient description is expected. From eq. (11) and the
expression of $C(\rho_{AB})$, one has
$$\left[C_a^{(AB)}
\right]^2=Tr\left(\rho_{AB}\tilde{\rho}_{AB}\right)+\frac{1}{2}\tau\left(\left\vert\psi\right\rangle_{ABC}\right),\eqno{(13)}$$and
$$C^2(\rho_{AB})=Tr\left(\rho_{AB}\tilde{\rho}_{AB}\right)-\frac{1}{2}\tau\left(\left\vert\psi\right\rangle_{ABC}\right).\eqno{(14)}$$
In particular, in terms of the twofold copy of
$\left\vert\psi\right\rangle_{ABC}$ (or $\rho_{AB}$), one can get
$$Tr\left(\rho_{AB}\tilde{\rho}_{AB}\right)=\sqrt{Tr[(\rho_{AB}\otimes\rho_{AB})\mathcal{B}]},\eqno{(14)}$$
where $\mathcal{B}=4P_-^{A_1A_2}\otimes P_-^{B_1B_2}$.
$Tr\left(\rho_{AB}\tilde{\rho}_{AB}\right)$ has been written in the
form of the expectation value of the self-adjoint operator
$\mathcal{B}$, hence it is measurable. Consider the other two pairs
of equations for $\rho_{AC}$ and $\rho_{BC}$ and the CKW equations,
one can determine all the entanglement, provided that two copies of
the reduced density matrix of two qubits are available. Thus a new
entanglement vector can be constructed by means of replacing
$E_{ii},E_{iii},E_{iv}$ by three
$Tr\left(\rho_{x}\tilde{\rho}_{x}\right)$ with $x$ denoting two
qubits. With the new entanglement vector, it not necessary to know
all the elements of the vector in order to determine a given
entanglement except the global entanglement.

We have considered that the measured quantum states are pure.
However, the imperfect preparation procedure may produce mixed
states. In practical experiment, analogous to Ref. [2], one can
discuss the deviation of measured entanglement by considering the
potential errors introduced by impure states and correct the
measurement values. Such an comparison procedure is quite simple and
omitted here.

 \emph{Summary.}- We have shown
that 3-tangle can be experimentally determined by a single
factorizable observable, provided that four copies of the state can
be provided, by which all the entanglement in terms of SLOCC
classification can be determined with the help of the measurable
bipartite concurrence or $Tr\rho_x\tilde{\rho}_x$. COA has also been
shown to be measurable by an interesting equation with explicit
physical meanings. We would like to emphasize that although reduced
density matrices of two qubits are mixed states, the exact
concurrence instead of the lower bound can be determined. Even
though four copies of the state are required, all the projective
measurements are only restrictive on the twofold copies, as has been
demonstrated recently in experiment. Furthermore, because a state
has to be prepared repeatedly in order to obtain reliable
measurement statistics in any experiment [2], a fourfold copy of a
state should be feasible in current experiment, which implies the
observation of all the entanglement of tripartite pure states of
qubits may be feasible.

\emph{Acknowledgement.}-This work was supported by the National
Natural Science Foundation of China, under Grant Nos. 10575017. The
author would like to thank F. X. Han for the revision of the paper.


\begin{thebibliography}{99}
\bibitem{[20]} Ref. [16] introduced two quantifications of entanglement for W-type states. For example the average residual
entanglement defined by
\begin{equation*}
\overline{E}\left(\left\vert\psi\right\rangle_{ABC}\right) =\frac{1}{3}%
\left(C^2(\rho_{AB})+C^2(\rho_{AC})+C^2(\rho_{BC})\right),
\end{equation*}
and the least entanglement of the reduced density matrices defined
by
\begin{equation*}
E_{min}\left(\left\vert\psi\right\rangle_{ABC}\right)=\min\left(C^2(%
\rho_{AB}),C^2(\rho_{AC}),C^2(\rho_{BC})\right).
\end{equation*}
Another alternative measure we propose can be given by
\begin{equation*}
E\left(\left\vert\psi\right\rangle_{ABC}\right)=\left(C^2(%
\rho_{AB})C^2(\rho_{AC})C^2(\rho_{BC})\right)^{1/3}.
\end{equation*}
\bibitem{[21]} One can determine COA by means of
solving the three monogamy equations presented in Ref. [18] with the
determined entanglement vector taken into consideration. However,
the same disadvantage mentioned latter can not be overcome.
\end{thebibliography}
\end{document}